\documentclass[aps, pra, reprint, superscriptaddress]{revtex4-1}
\usepackage{amssymb}
\usepackage{amsmath}
\usepackage{graphicx}
\usepackage{dcolumn}
\usepackage{bm}

\input{tcilatex}

\begin{document}

\title{Signatures of a non-thermal metastable state in copropagating quantum
Hall edge channels}
\author{Kosuke Itoh}
\author{Ryo Nakazawa}
\author{Tomoaki Ota}
\affiliation{Department of Physics, Tokyo Institute of Technology, 2-12-1 Ookayama,
Meguro, Tokyo, 152-8551, Japan.}
\author{Masayuki Hashisaka}
\affiliation{Department of Physics, Tokyo Institute of Technology, 2-12-1 Ookayama,
Meguro, Tokyo, 152-8551, Japan.}
\affiliation{NTT Basic Research Laboratories, NTT Corporation, 3-1 Morinosato-Wakamiya,
Atsugi, 243-0198, Japan.}
\author{Koji Muraki}
\affiliation{NTT Basic Research Laboratories, NTT Corporation, 3-1 Morinosato-Wakamiya,
Atsugi, 243-0198, Japan.}
\author{Toshimasa Fujisawa}
\email{fujisawa@phys.titech.ac.jp}
\affiliation{Department of Physics, Tokyo Institute of Technology, 2-12-1 Ookayama,
Meguro, Tokyo, 152-8551, Japan.}

\begin{abstract}
A Tomonaga-Luttinger (TL) liquid is known as an integrable system, in which
a non-equilibrium many-body state survives without relaxing to a thermalized
state. This intriguing characteristic is tested experimentally in
copropagating quantum Hall edge channels at bulk filling factor $\nu $ = 2.
The unidirectional transport allows us to investigate the time evolution by
measuring the spatial evolution of the electronic states. The initial state
is prepared with a biased quantum point contact, and its spatial evolution
is measured with a quantum-dot energy spectrometer. We find strong evidences
for a non-thermal metastable state in agreement with the TL theory before
the system relaxes to thermal equilibrium with coupling to the environment.
\end{abstract}

\date{\today }
\maketitle

Electron-electron interaction in usual conductors is often considered to
bring the system in a thermalized state irrespective of the initial states 
\cite{BookGemmer,BookAltshulerAronovEE,PothierThermalize}. In the case of
one-dimensional Tomonaga-Luttinger (TL) liquids with interacting electrons,
the integrable TL model suggests the presence of many conserved quantities
and the absence of thermalization processes \cite%
{GiamarchiBook,GutmanPRL2008,Iucci2009}. In the presence of weak
non-integrable interactions, the system exhibits two-stage equilibration
from an initial non-equilibrium state through an intermediate non-thermal
metastable state to a thermalized state \cite%
{PolkovnikovRMP2011,LevkivskyiPRB2012}. While such intriguing dynamics have
been observed in ultracold atoms \cite{KinoshitaNature2006,GringScience2012}%
, solid-state realization would open vast non-equilibrium many-body physics
particularly for transporting massive information. Edge channels in the
integer quantum Hall regime can host a chiral TL liquid particularly at bulk
Landau filling factor $\nu =2$ with spin-up and -down edge channels \cite%
{BookEzawa,ChangRMB}. Electrons in the channels are mutually interacting,
and collective excitation (plasmon) modes appear as the charge and spin (or
dipole) modes, which have symmetric and anti-symmetric charge distribution,
respectively, for the two channels if drift velocity difference is
negligible \cite{BergPRL2009}. This spin-charge separation has been
identified in various measurements such as time- and spin-resolved
measurement \cite{FreulonNatComm-SC,HashisakaNatPhys}, frequency-domain
plasmon interference \cite{BocquillonNatComm2013}, and shot-noise detection 
\cite{InouePRL2014}. A promising scheme for studying the equilibration
dynamics is quantum dot (QD) energy spectroscopy for a non-equilibrium state
prepared by a biased quantum point contact (QPC) \cite%
{AltimirasNatPhys2010,LevkivskyiPRB2012}, where non-thermal states can be
identified by observing non-Fermi distribution functions. Although the
previous work \cite{leSueurPRL2010} was successful in observing a spectral
change, the non-thermal metastable state was not resolved as the resulting
spectrum looked like a Fermi distribution function. As this can be explained
by either the TL model \cite{KovrizhinPRL2012} or a stochastic scattering
model \cite{LundePRB2010}, conclusive evidence is highly desirable.

In this work, we used the same QD-QPC scheme but investigated systematically
to see how the energy distribution function changes with the initial state
and the traveling distance. The expected non-thermal metastable state is
successfully identified with an arctangent distribution function by setting
the QPC at a low tunneling probability. The spectral change in the first
equilibration is consistent with the plasmon excitations based on the TL
model. The second equilibration toward cold Fermi distribution suggests weak
coupling to the environment. In this way, the edge channels provide a unique
opportunity for studying the integrability in a solid state system.

Figure 1(a) shows a schematic measurement setup for investigating
copropagating edge channels, C$_{\uparrow }$ for spin up and C$_{\downarrow
} $ for spin down, along a side of a two-dimensional electron system (2DES)
at $\nu $ = 2 under a perpendicular magnetic field $B$. The two ohmic
contacts on both ends are always grounded at base temperature $T_{\mathrm{%
base}}$. Non-equilibrium charge is injected from similar $\nu =2$ edge
channels, shown in the lower left, with bias voltage $V_{\mathrm{S}}$
through a QPC at conductance $(e^{2}/h)D$. Here, channel C$_{\uparrow }$ can
be excited by spin-up tunneling at $0<D<1$, and C$_{\downarrow }$ by
spin-down tunneling at $1<D<2$. The charge flows to the downstream, and the
electronic state at the distance $L$ from the QPC is investigated by a QD
spectrometer with an energy level $\varepsilon $ that can be tuned with the
QD gate voltage $V_{\mathrm{QD}}$. With appropriate bias voltage $V_{\mathrm{%
D}}$ on similar $\nu =2$ edge channels in the lower right, the current $I_{%
\mathrm{D}}(\varepsilon )$ through the dot level can be made proportional to
the energy distribution function $f_{\uparrow }\left( \varepsilon \right) $
in channel C$_{\uparrow }$ as described below.

\begin{figure}[tbp]
\begin{center}
\includegraphics[width = 3.15in]{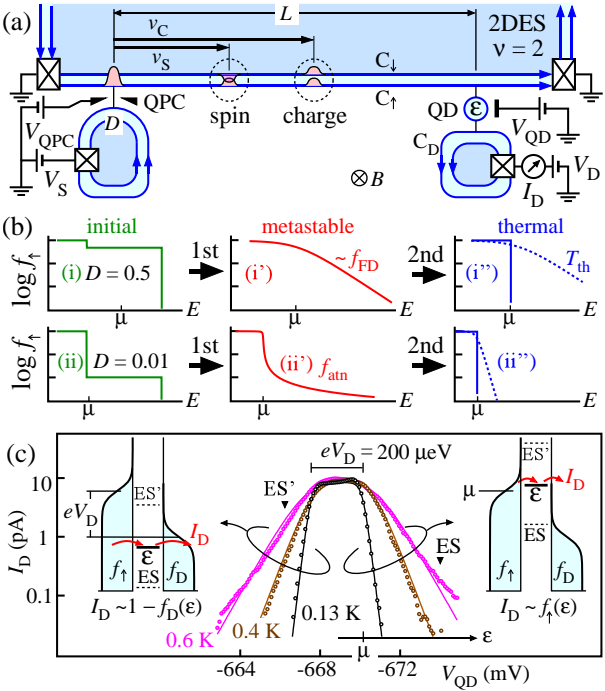}
\end{center}
\caption{(a) Schematic setup for the energy spectroscopy on copropagating
channels C$_{\uparrow }$ and C$_{\downarrow }$. (b) Expected two-stage
equilibration from initial states with double-step distribution function in
panel (i) at $D=0.5$ and (ii) at $D=0.01$, through metastable states in (i')
and (ii'), to thermalized states in (i\textquotedblright ) and
(ii\textquotedblright ) with dashed lines for a closed system and solid
lines for an open system. (c) Current profiles of the QD spectrometer at
various lattice temperatures. The energy diagrams in the left and right
insets show that the currents on the left and right sides are proportional
to the distribution functions $1-f_{_{\mathrm{D}}}$ and $f_{\uparrow }$,
respectively. Current through ES (ES') is allowed only when the ground state
at $\protect\varepsilon $ is occupied (empty).}
\label{FIG1}
\end{figure}

The initial energy distribution function in C$_{\uparrow }$ is expected to
be a double-step function with height $D$ and width $eV_{\mathrm{S}}$ by
assuming energy-independent tunneling probability, as shown in panel (i) for 
$D=0.5$ and (ii) for $D=0.01$ of Fig. 1(b). For $D<1$, one may consider that
single electrons are randomly injected to C$_{\uparrow }$ at a rate of $%
(e/h)DV_{\mathrm{S}}$. The uncertainty relation implies that each electron
wavepacket has a spread $h/eV_{\mathrm{S}}$ in time, and $vh/eV_{\mathrm{S}}$
in space for velocity $v$ in the channel. The coupling between C$_{\uparrow
} $ and C$_{\downarrow }$ splits each wavepacket into charge and spin
wavepackets as illustrated in Fig. 1(a) \cite{HashisakaNatPhys}. The length
required for the spin-charge separation is given by $\ell _{\mathrm{SC}}=hv_{%
\mathrm{SC}}/eV_{\mathrm{S}}$ with the relative velocity $v_{\mathrm{SC}}=v_{%
\mathrm{C}}v_{\mathrm{S}}/\left( v_{\mathrm{C}}-v_{\mathrm{S}}\right) $ for
charge and spin velocities $v_{\mathrm{C}}$ and $v_{\mathrm{S}}$,
respectively. Levkivskyi and Sukhorukov have calculated the energy
distribution function at large distances beyond the spin-charge separation
length $\ell _{\mathrm{SC}}$ \cite{LevkivskyiPRB2012}. It is close to, but
should be slightly different from, the Fermi distribution function $f_{%
\mathrm{FD}}(E)=[1+\mathrm{e}^{\left( E-\mu \right) /k_{\mathrm{B}}T_{%
\mathrm{th}}}]^{-1}$ at thermalization temperature $T_{\mathrm{th}}=\sqrt{%
\frac{3}{2}}\frac{1}{\pi k_{\mathrm{B}}}\sqrt{D\left( 1-D\right) }eV_{%
\mathrm{S}}$ when the tunneling is frequent at $D\simeq 0.5$ [panel (i') in
Fig. 1(b) for $D=0.5$]. Here, $\mu $ is the corresponding chemical
potential. In contrast, when the tunneling events are sparse ($D\ll 1$), a
non-thermal metastable state with a non-trivial distribution function of an
arctangent form $f_{\mathrm{atn}}\left( E\right) =\frac{1}{2}-\arctan \left(
E/\Gamma \right) /\pi $ (a Lorentzian function in $df/dE$ \cite%
{LevkivskyiPRB2012}) is expected to emerge with $\Gamma =2eDV_{\mathrm{S}%
}/\pi $ [panel (ii') for $D=0.01$]. Intriguingly, no scattering happens even
when a fast charge wavepacket overtakes a slow spin wavepacket, and thus
there should be no further thermalization processes in the integrable model.
Actual devices may have other thermalization processes. If the
thermalization is associated with the non-integrable interaction within the
channel, the system may relax to a heat-conserved thermalized state with a
Fermi distribution function at $T_{\mathrm{th}}$ [the dashed lines in panels
(i\textquotedblright ) and (ii\textquotedblright )] after a long travel. If
the system is weakly coupled to the environment, the system relaxes to a
thermalized state at $T_{\mathrm{base}}$ (the solid lines). We shall
investigate such two-stage equilibration.

We used a couple of devices with different length $L$ ranging from 0.12 to
15 $\mu $m between the QD and QPC (See Appendix A). They are fabricated in
standard AlGaAs/GaAs heterostructures with the electron density of 2.9 and
3.1 $\times $10$^{11}$ cm$^{-2}$ and low-temperature mobility of 1.6 and 1.9 
$\times $10$^{6}$ cm$^{2}$/Vs, and measured at $B$ = 6 and 7.5 T,
respectively for $L\geq $ 5 and $L\leq $ 0.5 $\mu $m devices in a dilution
refrigerator at $T_{\mathrm{base}}$ = 80 - 110 mK. Finite current in $I_{%
\mathrm{D}}$ is observed when the energy level $\varepsilon $ of the ground
state is located in the transport window of the width $eV_{\mathrm{D}}$
(typically 200 $\mu $eV), as shown in Fig. 1(c). The temperature dependence
shows a clear heating effect in both sides of the peak. All traces are
fitted nicely with the Fermi distribution function over more than two orders
of magnitude. As shown in the insets, $I_{\mathrm{D}}$ is proportional to $%
f_{\uparrow }$ on the right side, where we focus in the following
measurements, and to $\left( 1-f_{\mathrm{D}}\right) $ with distribution
function $f_{\mathrm{D}}$ in the drain channel C$_{\mathrm{D}}$ on the left
side. In practice, excited states in the QD may contribute additional
current (See Appendix B). The downward triangles labeled ES and ES' in all plots
represent the conditions of excited states being aligned to one of the
chemical potentials in the channels. Data in Fig. 1(c) shows that the
excited states play a minor role when the channels show Fermi distribution
functions.

Now, we investigate non-equilibrium states with the QPC. First, we focus on
the first equilibration occurring at $L\sim \ell _{\mathrm{SC}}$. Since $%
\ell _{\mathrm{SC}}$ is tunable with $V_{\mathrm{S}}$, the first
equilibration can be studied with varying $L$ and $\ell _{\mathrm{SC}}$.
Here, $\ell _{\mathrm{SC}}$ is estimated by using $v_{\mathrm{SC}}=27$ km/s
obtained in the following analysis. As summarized in Figs. 2(a) and 2(b),
the double-step current profile is clearly resolved in trace (i) taken at $L$
= 0.12 $\mu $m $\ll \ell _{\mathrm{SC}}=$ 1.1 $\mu $m ($V_{\mathrm{S}}=$ 100 
$\mu $V), but gradually smeared out as seen in trace (ii) at $L=$ 0.5 $\mu $%
m $<\ell _{\mathrm{SC}}=$ 1.1 $\mu $m and (iii) at $L=$ 0.5 $\mu $m $\sim
\ell _{\mathrm{SC}}=$ 0.7 $\mu $m ($V_{\mathrm{S}}=$ 150 $\mu $V). Their
step positions were determined from the peak positions in the derivative
(lower traces). The distance between the two steps, $\Delta $ evaluated in
energy, deviates from the original step width $eV_{\mathrm{S}}$ with
increasing $L$ and $V_{\mathrm{S}}$.

\begin{figure}[tbp]
\begin{center}
\includegraphics[width = 3.15in]{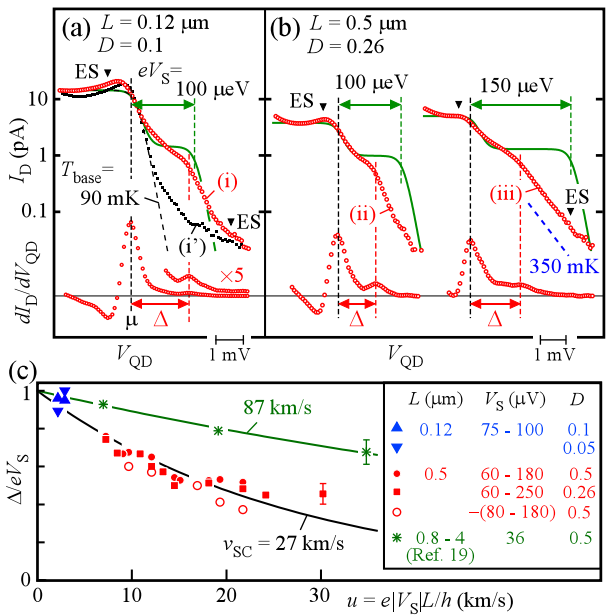}
\end{center}
\caption{(a) and (b) QD current profiles and their derivatives on the right
side of the current peak. The double-step feature is clear in (i), but
smeared out with reduced step distance $\Delta $ in (ii) at longer $L$ and
in (iii) at larger $V_{\mathrm{S}}$. The reference trace (i') for the
background excitation level is taken under opposite chirality. The black
dashed curve in (a) shows $f_{\mathrm{FD}}$ at $T_{\mathrm{base}}$ = 90 mK.
The green solid lines show the initial double-step function at $T_{\mathrm{%
base}}$. (c) The normalized step distance $\Delta /eV_{\mathrm{S}}$ as a
function of $u\equiv e\left\vert V_{\mathrm{S}}\right\vert L/h$. The solid
lines are exponential fits to our data with $v_{\mathrm{SC}}=$ 27 km/s and
the data in ref. \protect\cite{leSueurPRL2010} with 87 km/s.}
\label{FIG2}
\end{figure}

Figure 2(c) summarizes the normalized step width $\Delta /eV_{\mathrm{S}}$
as a function of the interaction strength defined by $u\equiv e\left\vert V_{%
\mathrm{S}}\right\vert L/h$. Data points taken at various $L$, $V_{\mathrm{S}%
}$, and $D$ follow single monotonic functions (solid lines). As we are not
aware of theoretical formula for this dependence, an exponential dependence $%
\Delta /eV_{\mathrm{S}}=\exp \left( -L/\ell _{\mathrm{SC}}\right) =\exp
\left( -u/v_{\mathrm{SC}}\right) $ is assumed for the solid lines with $v_{%
\mathrm{SC}}$ = 27 km/s for our devices and 87 km/s for the devices in Ref. 
\cite{leSueurPRL2010} with an additional surface gate. These values are
close to $v_{\mathrm{SC}}=$ 60 - 75 km/s obtained from a time-of-flight
experiment in Ref. \cite{HashisakaNatPhys}. The variation may stem from the
different geometries of the metal gate that partially screens the
interaction \cite{KumadaPRB2011,ProkudinaPRL2014,NoteVSC}. Note that the
observed $V_{\mathrm{S}}$ dependence in Fig. 2(c) does not agree with the 
\textit{stochastic} electron-electron scattering approach \cite{LundePRB-EE}%
, where the energy loss ($eV_{\mathrm{S}}-\Delta $) is found to be
independent of $V_{\mathrm{S}}$. Our systematic study supports that the
first-stage equilibration is associated with the \textit{deterministic}
spin-charge separation.

Let us investigate the energy distribution function at $L>\ell _{\mathrm{SC}%
} $. Figure 3(a) shows the current profile taken at $L=$ 5 $\mu $m and $%
D=0.005 $, where unusual distribution functions with a long tail appear. We
find excellent agreement with the theoretically predicted arctangent
function (the red solid lines), where the single parameter $\Gamma =2hI_{%
\mathrm{S}}/\pi e$ was determined from the measurement of $I_{\mathrm{S}}$.
As $\ell _{\mathrm{SC}}$ decreases with increasing $V_{\mathrm{S}}$, no
significant departure from the arctangent from is seen even at the longest
relative distance $L/\ell _{\mathrm{SC}}$ reaching 28 at $V_{\mathrm{S}}=$
600 $\mu $V ($\ell _{\mathrm{SC}}=$ 0.18 $\mu $m).

\begin{figure}[tbp]
\begin{center}
\includegraphics[width = 3.15in]{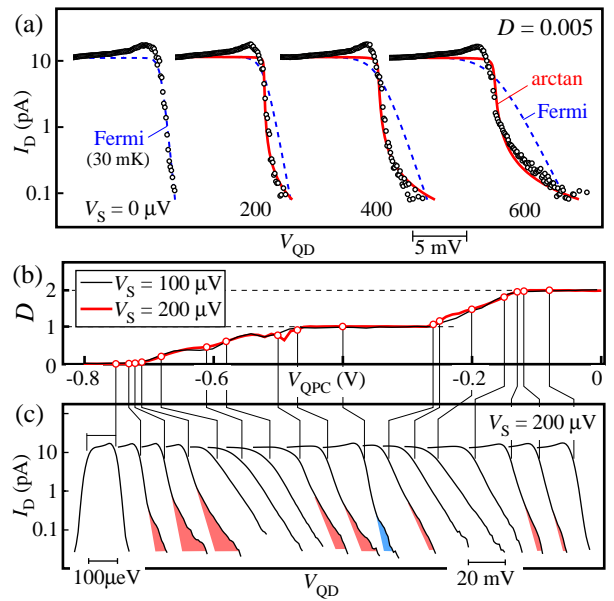}
\end{center}
\caption{(a) $V_{\mathrm{S}}$ dependence of the QD current profile obtained
at small $D$ = 0.005, showing an excellent agreement with the arctangent
function (the red solid lines). No features approaching to the Fermi
distribution function (the blue dashed lines for $T_{\mathrm{th}}$) are
seen. A peak near the Fermi edge is associated with the Fermi edge
singularity (See Appendix A3). (b) Gate voltage $V_{\mathrm{QPC}}$
dependence of dimensionless conductance $D$ of the QPC, where the series
resistance in the setup was subtracted. (c) QD current profiles at various $%
D $ marked by open circles in (b). Non-thermal current tail in (c) is
highlighted by red and blue regions. The small tail in the blue region for $%
D=1$ might be induced by spin-flip tunneling between the channels. Each
profile in (a) and (c) is offset horizontally for clarity. The width of the
current peak at $D=0$ corresponds $eV_{\mathrm{D}}=$ 200 $\protect\mu $eV in
(a) and 100 $\protect\mu $eV in (c).}
\label{FIG3}
\end{figure}

To identify the region where the non-thermal state emerges, the QD current
spectra in Fig. 3(c) are taken at various $D$ ranging from 0 to 2 marked by
red circles in the QPC conductance steps of Fig. 3(b). While nearly Fermi
distribution [showing a straight line in the low-current region of Fig.
3(c)] appears under frequent-tunneling conditions $D\simeq 0.5$ and $D\simeq
1.5$, non-Fermi distribution with a tail (marked by red regions) are
observed under sparse-tunneling conditions $0<D<0.3$, $0.7<D<1.3$, and $%
1.7<D<2$. Quantitatively similar behavior is observed for the spin-up ($%
0<D<1 $) and spin-down ($1<D<2$) tunneling, consistent with the
interpretation that the energy exchange occurs via spin-charge separation.
Detailed analysis on this data is shown in Appendix C.

If the current tail is associated with the non-thermal state of the TL
model, it should be stable for long traveling before the second
equilibration comes in. As shown in Fig. 4(a), non-thermal states showing
non-exponential current tails are well developed at $L=$ 0.5 $\mu $m close
to $\ell _{\mathrm{SC}}=$ 0.55 $\mu $m ($V_{\mathrm{S}}=$ 200 $\mu $V) and
greater than $\ell _{\mathrm{SC}}=$ 0.28 $\mu $m ($V_{\mathrm{S}}=$ 400 $\mu 
$V). Similar current profiles are also seen at much longer distances $%
L\simeq $ 5 and 15 $\mu $m as compared to $\ell _{\mathrm{SC}}=$ 0.22 $\mu $%
m ($V_{\mathrm{S}}=$ 500 $\mu $V) in Fig. 4(b). Although they were measured
at slightly different conditions with different samples, quite similar
current profiles showing a long tail are reproduced in the wide range of $L$%
. This manifests the long-lived nature of the non-thermal state.

The data in Fig. 4(b) were measured using the same QD spectrometer while the
excitation being done with different QPCs. Although the QPC characteristics
are slightly different, the long tail of the distribution function seems to
be attenuated as $L$ increases from 5 to 15 $\mu $m. Similar attenuation
with the decay length of about $\ell _{\mathrm{leak}}=$ 20 $\mu $m is also
seen at $D=0.5$ and $V_{\mathrm{S}}=$ 200 $\mu $V (See Appendix C3). 
This suggests
heat leakage to the environment. Although we need further studies to
identify its origin, it could be related to spin-flip tunneling between the
channels \cite{KomiyamaPRBspinflip,MullerPRBspinflip}, plasmon scattering
with counterpropagating channels \cite%
{KamataNatNano2014,ProkudinaPRL2014,WashioPRB}, excitation in remote
channels \cite{HashisakaPRB2013}, and coupling to phonons \cite%
{TelangPRBphonon,ProkudinaPRB2010}.

\begin{figure}[tbp]
\begin{center}
\includegraphics[width = 3.15in]{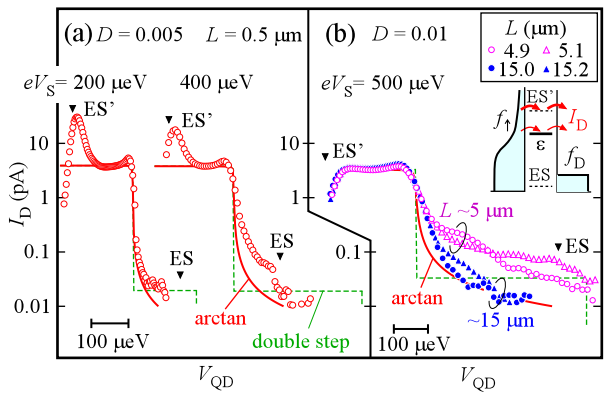}
\end{center}
\caption{(a) QD current profiles at $L\simeq $ 0.5 $\protect\mu $m. (b) QD
current profiles at $L\simeq $ 5 and 15 $\protect\mu $m. The unusual current
tail is observed all traces. The tail is larger than expected (the solid
line for the arctangent function) due to the excess current through ES' as
shown in the inset. The dashed lines show the double-step function at $T_{%
\mathrm{base}}$ = 0. The inset shows the energy diagram for QD spectroscopy
with excited states, ES and ES'.}
\label{FIG4}
\end{figure}

In this way, the expected non-thermal metastable state is clearly
demonstrated. The evidence is reinforced by the following arguments
excluding experimental artifacts.

We performed similar experiments at the reversed magnetic field where, due
to opposite chirality, plasmon excitations cannot reach the QD detector. We
observed no measurable influence on $I_{\mathrm{D}}$ at $L\geq $ 0.5 $\mu $%
m. A small excess current for the shortest distance of $L$ = 0.12 $\mu $m
[trace (i') at $B=$ -7.5 T in Fig. 2(a)] is possibly due to photon assisted
tunneling via electrostatic coupling between the QPC and the QD \cite%
{OnacPRL2006}. This ensures that the current tail observed at $B>$ 0 is
associated with the chiral plasmons in the edge channel.

Spin-flip tunneling between C$_{\uparrow }$ and C$_{\downarrow }$ can
influence the distribution function. This effect should be maximized at $D=1$%
, where we observed in a separate measurement with the same sample that
about 0.5 \% of injected electrons in C$_{\uparrow }$ experience tunneling
to C$_{\downarrow }$ during the propagation of 15 $\mu $m at $V_{\mathrm{S}%
}= $ 200 $\mu $V \cite{KomiyamaPRBspinflip,MullerPRBspinflip}. This unwanted
excitation might be the reason for having a small current tail (the blue
region) at $D=1$ in Fig. 3(c). The spin-flip tunneling should play a minor
role at $D\ll 1$.

The QPCs used in this work show nonlinear current-voltage characteristics 
\cite{QPC-QD-Yacoby2012} (See Appendix A2). 
The QPC used for Fig. 3 shows reasonable
linearity up to 200 $\mu $V, where the data in Fig. 3(c) was taken, as shown
by the small difference between the QPC conductance steps at $V_{\mathrm{S}%
}= $ 100 and 200 $\mu $V in Fig. 3(b). The linearity holds even up to $V_{%
\mathrm{S}}\simeq $ 500 $\mu $V for small $D=$ 0.005, where the clear
arctangent profile in Fig. 3(a) is obtained. Therefore, the nonlinearity
should play a minor role in the appearance of non-thermal states. However,
the nonlinearity can deflect or increase the distribution function from the
expected arctangent form, which could be the case for the data of Fig. 4.

While the calculated current profiles presented here were obtained by
solving rate equations with the ground state of the QD only, we also checked
the effects of excited states \cite{QD-ES-Fujisawa} (See Appendix B). For Fermi
distribution functions, the inclusion of the excited states in the
simulation did not change the current profile significantly. In contrast,
for arctangent functions, the current profile changed considerably when the
excited states have larger tunneling rates than the ground state. This also
explains why the observed current in Fig. 4 is greater than the calculated
one with the ground state only. We note that in Fig. 3 the measured current
in the tail agrees well with the calculation, where this QD shows small
tunneling rates for the excited states (See Appendix A3).

In summary, we have successfully observed non-thermal states with arctangent
energy distribution functions. This suggests that the system can be regarded
as an effectively closed quantum many-body system for a limited length ($%
<\ell _{\mathrm{leak}}$), despite the fact that ohmic contacts and the
measurement apparatus are attached. This would open the way to exploring
many-body quantum dynamics in the solid states \cite{CalzonaPRB2017-Entangle}%
.

\appendix
\renewcommand{\figurename}{FIG. S}
\setcounter{figure}{0}

\section{Device characteristics}

\subsection{Sample layout}

We used two quantum Hall samples, Sample A for $L$ = 0.15 and 0.5 $\mu $m
and Sample B for $L$ = 5 and 15 $\mu $m, fabricated in standard
modulation-doped AlGaAs/GaAs heterostructures.

For Sample A, the heterostructure shows electron density of 3.1 $\times $10$%
^{11}$ cm$^{-2}$ and low-temperature mobility of 1.9 $\times $10$^{6}$ cm$%
^{2}$/Vs. A scanning electron micrograph (SEM) of a test sample is shown in
Fig. S1(a), where surface metal gates with numbers 1 - 7 and the locations
of ohmic contacts ($\Omega _{1}$ - $\Omega _{4}$ and crossed boxes) are
shown.\ Edge channels C$_{\uparrow }$ and C$_{\downarrow }$ (not
illustrated) were formed along the lower side of gate 1 and 2 with
appropriate gate voltages under magnetic field $\left\vert B_{\mathrm{exp}%
}\right\vert =$ 7.5 T. A quantum dot (QD) marked by the white circle was
formed with gate 1, 2, 3, 4, and 5. Tunneling between C$_{\uparrow }$ and QD
was controlled with voltages on gate 1 and 2, and tunneling to the drain
channel toward ohmic contact $\Omega _{1}$ is tuned with the voltage on gate
5. For the measurement at $L$ = 0.12 $\mu $m [Fig. 2(a)],
the injector QPC was formed with gate 6. The trace (i) in Fig. 2(a) was
taken at positive $B_{\mathrm{exp}}$ with chirality from the QPC (gate 6) to
the QD, while trace (i') was taken at negative $B_{\mathrm{exp}}$ under
opposite chirality. The measurement at $L$ = 0.5 $\mu $m [Fig. 2(b)]
was performed with the QPC formed by gate 7 (and 1) under
negative $B_{\mathrm{exp}}$ with chirality from the QPC (7) to the QD.

For Sample B, the heterostructure shows electron density of 2.9 $\times $10$%
^{11}$ cm$^{-2}$ and low-temperature mobility of 1.6 $\times $10$^{6}$ cm$%
^{2}$/Vs. An SEM of a test sample is shown in Fig. S1(b). Edge channels C$%
_{\uparrow }$ and C$_{\downarrow }$ (not illustrated) were formed along the
upper side of the long gate 1 at $\left\vert B_{\mathrm{exp}}\right\vert =$
6 T. A QD spectrometer (white circle in the inset) was formed with gate 1,
5, 6, and 7, and a QPC injector is formed with gate 4 (and 1) for the
measurement at $L$ = 5 $\mu $m under positive $B_{\mathrm{exp}}$ with
chirality from the QPC(4) to the QD. All data in Fig. 3
was taken in this configuration. The measurement at $L$ = 5 and 15 $\mu $m
in Fig. 4(b) was performed with a QD spectrometer formed
with gate 1, 2, 3, and 4, and various QPCs with gate 5, 7, 8, and 9 under
negative $B_{\mathrm{exp}}$ with chirality from the QPCs to the QD. This
configuration with QPC(gate 8) is schematically shown in Fig. 1(a).

\begin{figure}[tbp]
\begin{center}
\includegraphics[width = 3.15in]{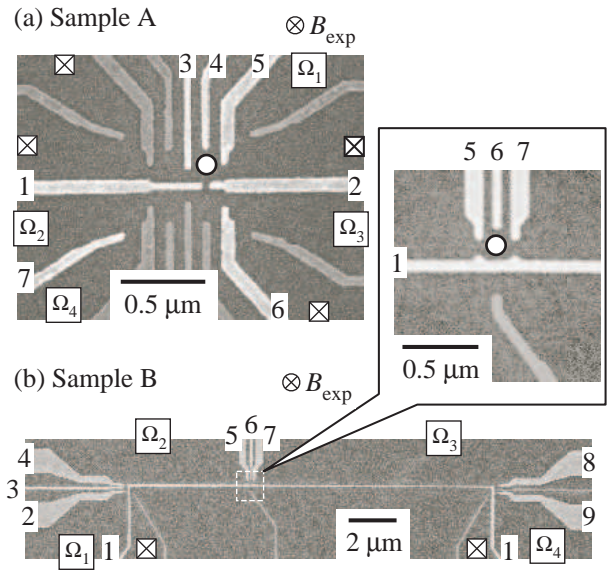}
\end{center}
\caption{(a) and (b) Scanning electron micrograph of Sample A in (a) and
Sample B in (b). The surface of the GaAs heterostructure appears dark.
Surface metal gates appear bright, but unused gates are darkened for better
visibility. }
\label{FIGS1}
\end{figure}

\subsection{QPC injector}

Figure S2(a) shows the quantized conductance of QPC(gate 4) in Sample B
which were used in the measurement of Fig. 3. The
dimensionless conductance $D$ was obtained from the total dc current $I_{%
\mathrm{S}}$ and the applied voltage $V_{\mathrm{S}}$ with the relation $V_{%
\mathrm{S}}=I_{\mathrm{S}}\left[ \left( h/e^{2}\right) /D+R_{\mathrm{ser}}%
\right] $. Here $R_{\mathrm{ser}}\sim $ 1 k$\Omega $ is the series
resistance including the ohmic resistances and the impedance of the
measurement system. The voltage drop on $R_{\mathrm{ser}}$ is less than 10
\% of $V_{\mathrm{S}}$ at $D=2$ and negligibly small at $D\ll 1$. The
conductance steps are clearly seen in Fig. S2(a) even at large bias voltage
of $V_{\mathrm{S}}$ = 400 $\mu $V. As the conductance traces at $V_{\mathrm{S%
}}$ = 100 $\mu $V and 200 $\mu $V are almost the same in the whole range $%
0<D<2$, the QPC is in a reasonably linear regime at $V_{\mathrm{S}}\leq $
200 $\mu $V. The QD current profiles in Fig. 3(c) were
taken under this condition. The nonlinear effect appears at higher bias as
seen in the deformed trace at $V_{\mathrm{S}}$ = 400 $\mu $V in Fig. S2(a).
Figure S2(b) shows the I-V characteristics of the QPC at low $D<0.1$. The
data shows quite linear conductance up to $V_{\mathrm{S}}$ = 500 $\mu $V for 
$D<0.02$. The arctangent current profile in Fig. 3(a) was
taken under this linear conductance region. In this way, the nonlinearity of
the QPC does not play a major role in the observation of non-Fermi
distribution functions.

When the QPC enters the nonlinear region, the current increases rapidly with
increasing $V_{\mathrm{S}}$ as shown in Fig. S2(b). As we evaluated $D\simeq
\left( h/e^{2}\right) I_{\mathrm{S}}/V_{\mathrm{S}}$ from the average
current $I_{\mathrm{S}}$, the total heat and the thermalization temperature $%
T_{\mathrm{th}}=\sqrt{\frac{3}{2}}\frac{1}{\pi k_{\mathrm{B}}}\sqrt{D\left(
1-D\right) }eV_{\mathrm{S}}$ is underestimated in the
nonlinear region. Moreover, the nonlinear effect can deflect the non-thermal
state away from the theoretically predicted state. For quantitative
analysis, we need further studies on the relation between electronic
excitation with a nonlinear QPC and the distribution function of the
non-thermal state.

\begin{figure}[tbp]
\begin{center}
\includegraphics[width = 3.15in]{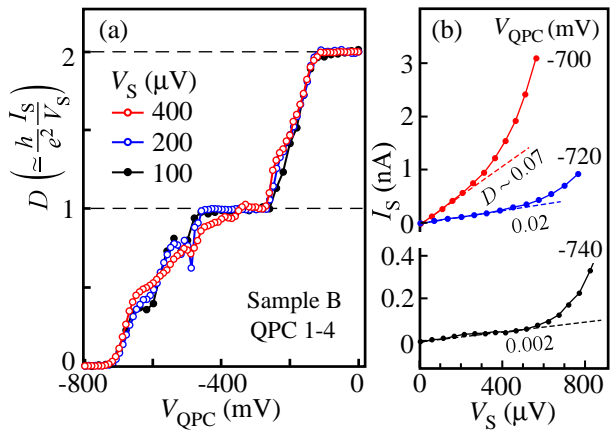}
\end{center}
\caption{(a) The dimensionless QPC conductance $D$ as a function of the gate
voltage $V_{\mathrm{QPC}}$. Nonlinearity of the QPC appears in some regions,
such as $-$650 mV $<V_{\mathrm{QPC}}<$ $-$350 mV at $V_{\mathrm{S}}=$ 400 $%
\protect\mu $V. (b) Current ($I_{\mathrm{S}}$) - voltage ($V_{\mathrm{S}}$)
characteristics of the QPC at several $V_{\mathrm{QPC}}$ for small $D$. The
low-bias linear dependence is shown by dashed lines.}
\label{FIGS2}
\end{figure}

\subsection{QD characteristics}

The QDs used in this work show the addition energy of 1 - 2 meV. The
particular Coulomb blockade peak used to measure the data in Fig. 3 
is shown in Fig. S3. The clear conductive region between the $N$-
and $(N+1)$-electron Coulomb blockade regions is seen with the onsets at $%
\varepsilon =\mu _{\mathrm{D}}$ and $\varepsilon =\mu _{\uparrow }$. In
addition to the transport through the ground state with $\varepsilon $,
several current steps associated with excited states are seen. The features
labeled ES (small steps at $V_{\mathrm{D}}<0$ and very faint steps marked by
dots at $V_{\mathrm{D}}>0$) are associated with the first excited state of $%
N $-electron QD (the level spacing $\Delta \sim $ 350 $\mu $eV). The
features labeled ES' and ES\textquotedblright\ are associated with the first
and second excited states of $(N+1)$-electron QD (the level spacing $\Delta
^{\prime }\sim $ 250 $\mu $eV between the ground state and the first excited
state). Extrapolating these excited-state features to the specific trace at $%
V_{\mathrm{D}}=$ 200 $\mu $V (the red trace) are marked by downward
triangles. All downward triangles shown in Fig. 1-4 were obtained in
this way.

The current shows a peak at $\varepsilon =\mu _{\uparrow }$, which can be
explained by the Fermi-edge singularity originated from the many-body effect
between the localized state in the QD and electrons in C$_{\uparrow }$ \cite%
{FermiEdge-EnsslinG2017,FermiEdge-Theory}. While an arctangent current
spectrum is suggested theoretically also for the Fermi-edge singularity, we
did not observe such spectrum in our samples [the trace for $D=0$ in Fig.
3(a)]. The Fermi-edge singularity should be significantly
diminished by smearing the Fermi edge with increasing the temperature \cite%
{FermiEdge-EnsslinG2017}. Therefore, we neglect the Fermi-edge singularity
in the analysis of current tail for the non-thermal state.

For proving the distribution function $f_{\uparrow }$ in C$_{\uparrow }$, it
is desirable to have smaller incoming tunneling rate $\Gamma _{\mathrm{%
\uparrow }}^{\left( \mathrm{G}\right) }$ as compared to the outgoing
tunneling rate $\Gamma _{\mathrm{D}}^{\left( \mathrm{G}\right) }$. We tuned
the gate voltages such that $\Gamma _{\mathrm{\uparrow }}^{\left( \mathrm{G}%
\right) }/\Gamma _{\mathrm{D}}^{\left( \mathrm{G}\right) }\sim 0.1$ for the
QD shown in Fig. S3. The asymmetric tunneling rates can be confirmed with
the prominent excited-state features running parallel to the current onset
at $\varepsilon =\mu _{\uparrow }$, as well as the Fermi-edge singularity
seen at $\varepsilon =\mu _{\uparrow }$ but not at $\varepsilon =\mu _{%
\mathrm{D}}$. We can deduce the tunneling rates from the saturated current
by neglecting the Fermi-edge singularity. For the QD shown in Fig. S3, the
tunneling rates across the left barrier to C$_{\uparrow }$ are $\Gamma _{%
\mathrm{\uparrow }}^{\left( \mathrm{G}\right) }\simeq $ 100 MHz for the
ground state, $\Gamma _{\mathrm{\uparrow }}^{\left( \mathrm{E}\right)
}\simeq $ 20 MHz for ES, and $\Gamma _{\mathrm{\uparrow }}^{\left( \mathrm{E}%
^{\prime }\right) }\simeq $ 40 MHz for ES'. As described below, small $%
\Gamma _{\mathrm{\uparrow }}^{\left( \mathrm{E}^{\prime }\right) }$ ($%
<\Gamma _{\mathrm{\uparrow }}^{\left( \mathrm{G}\right) }$) is desirable for
studying non-thermal states.

\begin{figure}[tbp]
\begin{center}
\includegraphics[width = 3.15in]{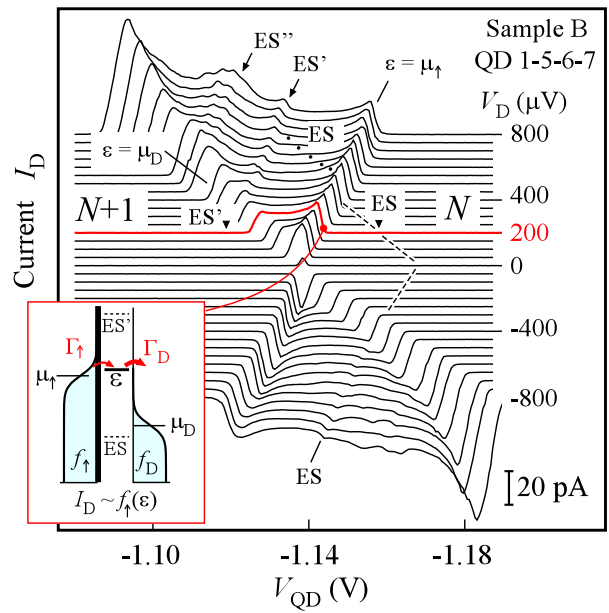}
\end{center}
\caption{The QD current $I_{\mathrm{D}}$ as a function of the gate voltage $%
V_{\mathrm{QD}}$ for various bias voltage $V_{\mathrm{D}}$, showing a
Coulomb peak between $N$- and $(N+1)$-electron Coulomb blockade regions.
Some excited-state features are marked. Each trace is offset for clarity.
The inset shows the energy diagram of the transport. }
\label{FIGS3}
\end{figure}

\section{QD spectroscopy}

Here, we characterize the QD spectroscopy particularly in the presence of
excited states in the QD. The QD current is calculated by using rate
equations for occupation probabilities in the QD states \cite{QD-ES-Fujisawa}%
. For simplicity, transport through the first excited state of $N$-electron
QD, ES, and the first excited state of $(N+1)$-electron QD, ES, as well as
the ground state are considered as shown in the energy diagram of Fig.
S4(a). The corresponding tunneling rates $\Gamma _{\uparrow \mathrm{,D}%
}^{\left( \mathrm{E}\right) }$, $\Gamma _{\uparrow \mathrm{,D}}^{\left( 
\mathrm{E}^{\prime }\right) }$, and $\Gamma _{\uparrow \mathrm{,D}}^{\left( 
\mathrm{G}\right) }$ are considered to be independent of the QD states but
asymmetric with respect to the left and right barriers; $\Gamma _{\mathrm{%
\uparrow }}^{\left( \mathrm{E}\right) }=\Gamma _{\mathrm{\uparrow }}^{\left( 
\mathrm{E}^{\prime }\right) }=\Gamma _{\mathrm{\uparrow }}^{\left( \mathrm{G}%
\right) }=$ 100 MHz and $\Gamma _{\mathrm{D}}^{\left( \mathrm{E}\right)
}=\Gamma _{\mathrm{D}}^{\left( \mathrm{E}^{\prime }\right) }=\Gamma _{%
\mathrm{D}}^{\left( \mathrm{G}\right) }=$ 1 GHz. Figure S4(b) shows the
calculated Coulomb peak with energy spacing $\Delta =$ 350 $\mu $eV for ES
and $\Delta ^{\prime }=$ 250 $\mu $eV for ES', where $\alpha V_{\mathrm{QD}}$
is the electrostatic potential of the QD. Here, both edge channels C$%
_{\uparrow }$ and C$_{\mathrm{D}}$ are considered to be in thermal
equilibrium at the base temperature\ of 80 mK ($k_{\mathrm{B}}T_{\uparrow
}=k_{\mathrm{B}}T_{\mathrm{D}}=$ 6.8 $\mu $eV). Some features associated
with ES and ES' are indicated by arrows and dots, and their extrapolations
to the $V_{\mathrm{D}}=$ 200 $\mu $V trace are marked by downward triangles.
The experimental features in Fig. S3 are well reproduced except for the step
heights (not adjusted) and the Fermi edge singularity.

\begin{figure}[tbp]
\begin{center}
\includegraphics[width = 3.15in]{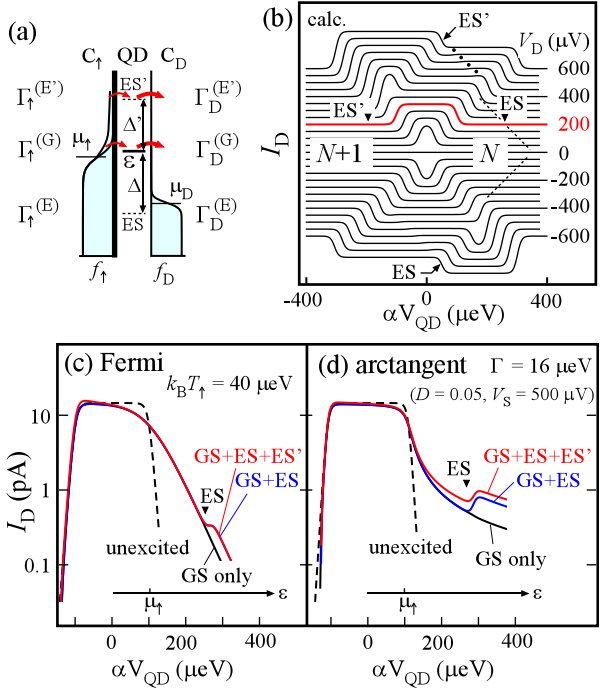}
\end{center}
\caption{(a) Energy diagram of transport through a QD. ES and ES' are first
excited states of $N$- and $(N+1)$-electron QDs, respectively. (b)
Calculated QD current $I_{\mathrm{D}}$ as a function of electrostatic
potential $\protect\alpha V_{\mathrm{QD}}$ at various $V_{\mathrm{D}}$. (c)
and (d) $I_{\mathrm{D}}$ at $V_{\mathrm{D}}=$ 200 $\protect\mu $V with a
heated Fermi distribution function for (c) and an arctangent function for
(d) in C$_{\uparrow }$. }
\label{FIGS4}
\end{figure}

The QD current profile changes when C$_{\uparrow }$ is excited. Solid lines
in Fig. S4(c) and S4(d) show the QD current calculated when Fermi
distribution function $f_{\mathrm{FD}}$ at $k_{\mathrm{B}}T_{\uparrow }=$ 40 
$\mu $eV in (c) and arctangent function $f_{\mathrm{atn}}$ (defined in the
main part) with $\Gamma =$ 16 $\mu $eV in (d) are considered for C$%
_{\uparrow }$. $\Gamma =$ 16 $\mu $eV corresponds to the case at $D=0.05$
and $V_{\mathrm{S}}=$ 500 $\mu $V. If there are no excited states in the QD,
the current (the black solid line labeled `GS only')\ is proportional to the
distribution function $f_{\uparrow }\left( \varepsilon \right) $ for $\alpha
V_{\mathrm{QD}}>0$. If ES is included in the calculation, the current (the
blue solid line labeled `GS+ES') exhibits a stepwise increase at $%
\varepsilon =\mu _{\mathrm{D}}+\Delta $ (marked by the downward triangle).
However, we did not observe clear signature of this current step in our
measurements, probably because energy relaxation rate from ES to GS is
larger than $\Gamma _{\mathrm{\uparrow }}^{\left( \mathrm{E}\right) }$ \cite%
{QD-ES-Fujisawa}.

If ES', as well as ES and GS, are considered in the calculation, the QD
current in Fig. S4(d) for the arctangent function is enhanced considerably
at $\varepsilon >\mu _{\uparrow }$ (the red solid line labeled
\textquotedblleft GS+ES+ES'\textquotedblright ). In contrast, no significant
change is seen in Fig. S4(c) with Fermi distribution function. The
difference can be understood with the energy diagram of Fig. 4(a), where the
high-energy tail of the arctangent (Fermi) distribution function $%
f_{\uparrow }$ gives non-negligible (negligible) current through ES'. The
rate-equation solution gives $I_{\mathrm{D}}\simeq \Gamma _{\mathrm{\uparrow 
}}^{\left( \mathrm{G}\right) }f_{\uparrow }\left( \varepsilon \right)
+\Gamma _{\mathrm{\uparrow }}^{\left( \mathrm{E}^{\prime }\right)
}f_{\uparrow }\left( \varepsilon +\Delta ^{\prime }\right) $ for $\Gamma _{%
\mathrm{\uparrow }}\ll \Gamma _{\mathrm{D}}$ and $\alpha V_{\mathrm{QD}}>0$,
which is independent of the energy relaxation rate from ES' to GS. The
second term is negligible for Fermi distribution functions, but can
contribute dominant current for the arctangent functions particularly when $%
\Gamma _{\mathrm{\uparrow }}^{\left( \mathrm{E}^{\prime }\right) }$ is
greater than $\Gamma _{\mathrm{\uparrow }}^{\left( \mathrm{G}\right) }$. For
the QD shown in Fig. S3, small contribution with $\Gamma _{\mathrm{\uparrow }%
}^{\left( \mathrm{E}^{\prime }\right) }\simeq 0.4\Gamma _{\mathrm{\uparrow }%
}^{\left( \mathrm{G}\right) }$ is expected. This explains why we see good
agreement with the expected arctangent profile at $D$ = 0.005 in Fig. 3(a).
However, larger contribution is suggested with $\Gamma _{%
\mathrm{L}}^{\left( \mathrm{E}^{\prime }\right) }\simeq 10\Gamma _{\mathrm{L}%
}^{\left( \mathrm{G}\right) }$ and $\Gamma _{\mathrm{L}}^{\left( \mathrm{E}%
^{\prime }\right) }\simeq 2\Gamma _{\mathrm{L}}^{\left( \mathrm{G}\right) }$
for the QDs used to take the data shown in Fig. 4(a) and 4(b), respectively.
This could be a reason for showing large current tail greater than the
predicted arctangent form.

\section{Non-thermal spectra}

\subsection{Transition between arctangent and Fermi functions}

Figure S5 shows a $D$ dependence of the QD current profile taken at $V_{%
\mathrm{S}}=$ 500 $\mu $V ($\ell _{\mathrm{SC}}=$ 0.22 $\mu $m) and $L=$ 5 $%
\mu $m. We find excellent agreement with the theoretically predicted
arctangent function (red solid lines) $f_{\mathrm{atn}}\left( E\right) =%
\frac{1}{2}-\arctan \left( E/\Gamma \right) /\pi $ at $D=$ 0.005 with a
single parameter $\Gamma =2eDV_{\mathrm{S}}/\pi $ that was determined from
the measurement of $I_{\mathrm{S}}$ with $D\simeq \left( h/e^{2}\right) I_{%
\mathrm{S}}/V_{\mathrm{S}}$, and with the Fermi distribution function (blue
dotted lines for $T_{\mathrm{th}}=\sqrt{\frac{3}{2}}\frac{1}{\pi k_{\mathrm{B%
}}}\sqrt{D\left( 1-D\right) }eV_{\mathrm{S}}$) at $D=$ 0.5. Smooth
transition between them is clearly resolved at $0.005<D<0.5$.

\begin{figure}[tbp]
\begin{center}
\includegraphics[width = 3.15in]{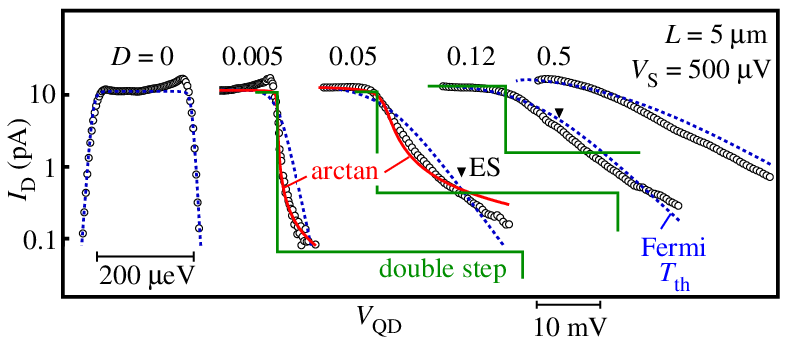}
\end{center}
\caption{$D$ dependence of the QD current at $V_{\mathrm{S}}=$ 500 $\protect%
\mu $V and $L=$ 5 $\protect\mu $m.}
\label{FIGS5}
\end{figure}

\subsection{$D$ dependence}

Here, we provide a detailed analysis of the $D$ dependence of QD current
profiles for $L=$ 5 $\mu $m presented in Figs. 3(b) and 3(c).
For convenience, they are also shown in Figs. S6(a) and S6(b). The
non-thermal distribution function can be characterized by non-exponential
current rolloff. To clarify the region showing non-thermal distribution
functions, we compare the slope of the rolloff at different $f$ values. For
convenience, we defined two temperatures, $T_{\mathrm{0.5}}$ deduced from
the data at $f\simeq 0.5$ [$k_{\mathrm{B}}T_{\mathrm{0.5}}=-\left\{ 2d(\ln
f)/dE|_{f=0.5}\right\} ^{-1}$], and $T_{\mathrm{0.01}}$ at $f\simeq 0.01$ [$%
k_{\mathrm{B}}T_{\mathrm{0.01}}=-\left\{ d(\ln f)/dE|_{f=0.01}\right\} ^{-1}$%
]. The factor 2 difference in the definitions is introduced such that Fermi
distribution functions always show $T_{\mathrm{0.5}}=T_{\mathrm{0.01}}$. As
shown in Figs. S6(c) and S6(d), while nearly Fermi distribution function ($%
T_{\mathrm{0.5}}\simeq T_{\mathrm{0.01}}$) appears at frequent-tunneling
conditions $D\simeq 0.5$ and $D\simeq 1.5$, non-Fermi distribution functions
with a tail ($T_{\mathrm{0.5}}<T_{\mathrm{0.01}}$) are observed at
sparse-tunneling conditions $0<D<0.3$, $0.7<D<1.3$, and $1.7<D<2$
(highlighted by vertical bars connecting circles for $T_{\mathrm{0.01}}$ and
squares for $T_{\mathrm{0.5}}$). Qualitatively the same feature for both
spin-up tunneling ($0<D<1$) and spin-down tunneling ($1<D<2$) is consistent
with the energy exchange associated with the spin-charge separation.

\begin{figure}[tbp]
\begin{center}
\includegraphics[width = 3.15in]{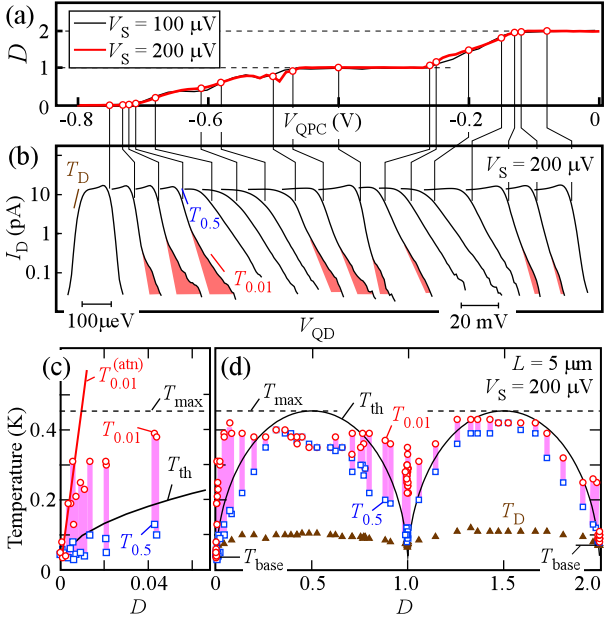}
\end{center}
\caption{(a) Gate voltage $V_{\mathrm{QPC}}$ dependence of dimensionless
conductance $D$. (b) The QD current spectra obtained at various $V_{\mathrm{%
QPC}}$\ marked by circles in (a). Each trace is offset horizontally for
clarity. Non-thermal distribution is highlighted by magenta regions. (c) and
(d) $D$ dependence of electron temperatures, $T_{0.5}$, $T_{0.01}$, and $T_{%
\mathrm{D}}$ obtained from the slopes depicted in (b), and $T_{0.01}^{(%
\mathrm{atn})}$, $T_{\mathrm{th}}$, and $T_{\mathrm{\max }}$ from the model.
A part of (c) is replotted in (b) with a magnified scale.}
\label{FIGS6}
\end{figure}

The theoretically derived arctangent formula suggests that $T_{\mathrm{0.01}%
} $ increases linearly with $D$, in a form $T_{f}^{\left( \mathrm{atn}%
\right) }=2eV_{\mathrm{S}}D/\pi ^{2}k_{\mathrm{B}}f$ for small $f$ (= 0.01
for the present case). Experimental plots of $T_{\mathrm{0.01}}$ are
scattered around $T_{f}^{\left( \mathrm{atn}\right) }$ only for small $%
D<0.01 $ (the solid line for $T_{0.01}^{\left( \mathrm{atn}\right) }$) in
Fig. S6(c). However, $T_{\mathrm{0.01}}$ does not follow $T_{f}^{\left( 
\mathrm{atn}\right) }$ at $D>0.1$, and saturates at $D>0.04$. It is almost
constant over the wide range of $D$ ($0.04<D<0.9$ and $1.2\lesssim D\lesssim
1.8$)\ as seen in Fig. S6(d). While non-thermal states are seen in the wide
range, the theoretically predicted arctangent form appear in the narrow
range $D<0.01$.

The constant slope ($T_{\mathrm{0.01}}$) can be understood with the Fourier
spectrum of the wavepackets. The frequency range of the packets is given by $%
eV_{\mathrm{S}}/h$, while the excitation amplitude is proportional to $\eta
=4D(1-D)$. If the excitation is characterized by the Fermi distribution
function $f_{\mathrm{FD}}(E;T_{\max })$ with the maximum temperature $%
T_{\max }$ of $T_{\mathrm{th}}$ at $D=0.5$ ($\eta =1$) as suggested by the
theory \cite{LevkivskyiPRB2012}, the corresponding plasmon distribution
follows the Bose distribution function $g_{\mathrm{BE}}\left( \omega
;T_{\max }\right) $ of plasmon frequency $\omega $ \cite%
{vonDelftAnnPhys1998,GutmanPRL2008}. When the excitation is reduced by the
factor $\eta $ ($<1$), the plasmon distribution would also be reduced to $%
\eta g_{\mathrm{BE}}\left( \omega ;T_{\max }\right) $, and the corresponding
electronic distribution can be written as $(1-\eta )f_{\mathrm{FD}}(E;T_{%
\mathrm{base}})+\eta f_{\mathrm{FD}}(E;T_{\max })$ by taking a crude
approximation in the fermionization \cite{WashioPRB}. Since the second term
dominantly determines $T_{\mathrm{0.01}}$ at $0.01<\eta \ll 1$, this
explains the appearance of almost constant $T_{\mathrm{0.01}}$ close to $%
T_{\max }$ [dashed lines in Figs. S6(c) and S6(d)] with the allowance of the
heat leakage (about 20\%) described below. This simple model does not work
at $\eta <0.01$, where the arctangent profile is well developed.

\subsection{Heat leakage}

\begin{figure}[tbp]
\begin{center}
\includegraphics[width = 3.15in]{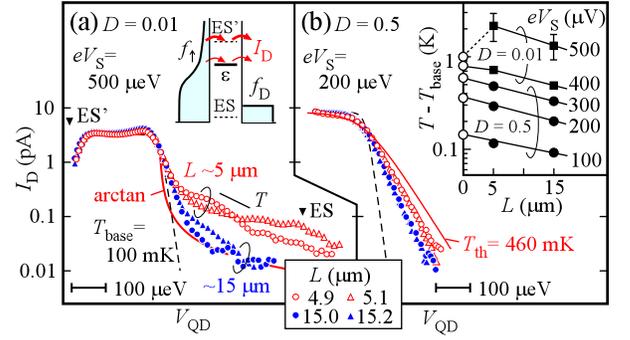}
\end{center}
\caption{(a) and (b) QD current spectra showing non-thermal states at $L$ =
5 and 15 $\protect\mu $m for (a) $D=$ 0.01 and (b) 0.5. The inset to (b)
shows the decay of the effective temperature $T$ obtained from the slope of
the tail.}
\label{FIGS7}
\end{figure}

The heat leakage is further investigated with the data shown in Fig. S7 for $%
L\simeq $ 5 and 15 $\mu $m. These data were measured using the same QD
spectrometer while the excitation being done with different QPCs. For $%
D=0.01 $ in Fig. S7(a), the long tail of the distribution function is
clearly seen at $L\simeq $ 15 $\mu $m as well as at $L\simeq $ 5 $\mu $m.
However, the current tail seems to be attenuated as $L$ increases from 5 to
15 $\mu $m, implying heat leakage to the environment. Since these QPCs enter
the nonlinear regime at $V_{\mathrm{S}}>$ 200 - 400 $\mu $V, ambiguity in
the injected heat is concerned. A similar signature of heat leakage is also
seen in the QD current profiles in Fig. S7(b) taken at $D=0.5$ and $V_{%
\mathrm{S}}=$ 200 $\mu $V in the reasonably linear conductance region. We
evaluated the effective temperature $T$ from the slope of the tail in Figs.
S7(a) and S7(b). As shown in the inset, $T-T_{\mathrm{base}}$ (solid
symbols), when plotted as a function of $L$, shows an exponential decay from 
$T_{\mathrm{th}}-T_{\mathrm{base}}$ (open circles) with no clear dependence
on $V_{\mathrm{S}}$ and $D$. This decay length $\ell _{\mathrm{leak}}\sim $
20 $\mu $m is much longer than the spin-charge separation length ($\ell _{%
\mathrm{SC}}=$ 0.2 - 1 $\mu $m at $V_{\mathrm{S}}=$ 500 - 100 $\mu $V by
assuming $v_{\mathrm{SC}}=$ 27 km/s), which manifests the long-lived
metastable nature of the non-thermal state.

\begin{acknowledgments}
We thank Yasuhiro Tokura and Keiji Saito for fruitful discussions. This work
was supported by JSPS KAKENHI (JP26247051, JP15H05854), International
Research Center for Nanoscience and Quantum Physics at Tokyo Institute of
Technology, and Nanotechnology Platform Program of MEXT.
\end{acknowledgments}


\begin{thebibliography}{99}
\bibitem{NoteVSC} $v_{\mathrm{SC}}$ is sensitive to $v_{\mathrm{S}}$ rather
than $v_{\mathrm{C}}$. The effect of the metallization on $v_{\mathrm{S}}$
is not understood well.
\end{thebibliography}
\end{document}